\documentclass[resetfootnote, twocolumn, tighten]{aastex701}

\usepackage[normalem]{ulem}
\usepackage{soul}

\usepackage{color}

\def\oiii{[O\,{\sc iii}]}

\def\hi{H\,{\sc i}}

\def\sii{[S\,{\sc ii}]}

\def\kms{\relax \ifmmode {\,\rm km\,s}^{-1}\else \,km\,s$^{-1}$\fi}
\newcommand{\Msun}{\ensuremath{\,\mathrm{M}_\odot}}
\newcommand{\Lsun}{\ensuremath{\,\mathrm{L}_\odot}}

\newcommand{\ltsimeq}{\raisebox{-0.6ex}{$\,\stackrel
        {\raisebox{-.2ex}{$\textstyle <$}}{\sim}\,$}}
\newcommand{\gtsimeq}{\raisebox{-0.6ex}{$\,\stackrel
        {\raisebox{-.2ex}{$\textstyle >$}}{\sim}\,$}}

\usepackage{ulem}
\usepackage{hyperref}
\usepackage{datetime2}

\received{2026 February 15}
\accepted{2026 March 12}

\begin{document}

\title{Detection of a molecular hydrogen envelope around nova GK Persei}

\author[orcid=0000-0002-9670-4824,sname='Banerjee']{D. P. K. Banerjee}
\affiliation{Astronomy and Astrophysics Division, Physical Research Laboratory, Ahmedabad, India 380009}
\email[show]{dpkb12345@gmail.com}  

\author[orcid=0000-0002-3142-8953,gname='Evans']{A.Evans} 
\affiliation{Astrophysics Research Centre, Lennard Jones Laboratories, Keele University, Keele, Staffordshire ST5 5BG, UK}
\email{a.evans@keele.ac.uk}

\author[orcid=0000-0003-2196-9091,gname='Liimets']{T.Liimets}
\affiliation{Tartu Observatory, University of Tartu, Observatooriumi 1, T\-{o}ravere 61602, Estonia}
\email{tiina.liimets@ut.ee}

\author[orcid=0000-0001-6567-627X,sname='Woodward']{C. E. Woodward}
\affiliation{Minnesota Institute for Astrophysics, School of
Physics \& Astronomy, 116 Church Street SE, University of Minnesota,
Minneapolis, MN 55455, USA}
\email[show]{chickw024@gmail.com}

\author[orcid=0000-0003-2824-3875,sname='Geballe']{T. R. Geballe}
\affiliation{Gemini Observatory/NSF's NOIRLab, 670 N. Aohoku Place, Hilo, HI, 96720, USA}
\email{tom.geballe@noirlab.edu}

\author[orcid=0000-0002-1457-4027,sname='Joshi']{V. Joshi}
\affiliation{Physical Research Laboratory, Navrangpura,  Ahmedabad, Gujarat 380009, India}
\email{onlyvishal@gmail.com}

\author[orcid=0000-0002-1359-6312,sname='Starrfield']{S. Starrfield}
\affiliation{School of Earth and Space Exploration, Arizona State University, Box 876004, Tempe, AZ 85287-6004, USA}
\email{sumner.starrfield@gmail.com}

\correspondingauthor{C.E. Woodward}
\collaboration{all}{}

\begin{abstract}
The eruption of Nova Persei 1901 (GK~Per) occurred 125 yrs ago; remarkably it still holds major surprises.
Using data from the Spectro-Photometer for the History of the Universe, Epoch of  Reionization, and 
Ices Explorer (SPHEREx), we find it has a bipolar molecular hydrogen shell. This shell,
which has dimensions $18\arcmin\times10\arcmin$, is co-spatial with the H$\alpha$ nebulosity surrounding
the nova, which is purported to be an ancient planetary nebula (PN). The shell is detected 
most strongly in the 0--0 $S$(9) 4.6947~$\mu$m line.  A filament of emission in the $S$(9) 4.6947~$\mu$m line 
is seen $\simeq 45\arcsec$\ South-West of GK~Per.  This coincides, over much of its length,
with the site of X-ray and non-thermal radio emission where the 1901 nova ejecta impinges on the ambient medium.
We propose that the H$_{2}$ emission from the filament arises from the predicted neutral zone 
between the forward and reverse shocks. Since it is common for bipolar PNe to be  accompanied by 
H$_{2}$ envelopes,  it ostensibly suggests that the $18\arcmin\times10\arcmin$ nebulosity is a 
conventional PN with a luminous, ionizing central source. We show this is not the case, and that the 
H$\alpha$ nebulosity may be surrounding gas belonging to pre-existing material that was ionized during the 1901 
eruption. The ionized gas is presently undergoing recombination on a timescale of $\simeq$ 3000 years, explaining why the 
nebulosity is still visible.
\end{abstract}


\section{Introduction} 
\label{intro}

Nova Persei 1901 (GK~Per) was the first bright classical nova of the 20th century. Seven months after 
eruption, superluminal light echoes were detected \citep{1901ApJ....14..167R,1901ApJ....14..293R,1902ApJ....15..129R}; 
however ejecta from the nova outburst became visible only 15 years later \citep{1916BHarO.621....1B}. 
Since then, the remnant of the 1901 outburst  has been extensively studied 
\citep[e.g.,][and references therein]{1989ApJ...344..805S,1993MNRAS.263..335A,2012ApJ...761...34L,2016A&A...595A..64H}. 
It has a round, overall morphology consisting of almost a thousand cometary-like structures, as 
revealed by high-resolution images from the Hubble Space Telescope \citep{2012AJ....143..143S}, and 
currently extends to a radius of $\sim1\arcmin$. The expansion velocities of individual knots in the 
remnant have been measured to mostly lie between 600 and 1000\kms\ by \cite{2012ApJ...761...34L}, 
who found that the remnant is best described as a thick spherical shell. \cite{2016A&A...595A..64H}, using
the same data, proposed an equally satisfactory alternative model, consisting of an equatorial barrel 
feature with polar cones.

A considerably larger ($\sim20\arcmin$) bipolar structure, compared to the nova remnant, was discovered 
at 60 \micron{} and 100 \micron{} in Infrared Astronomical Satellite \citep[IRAS,][]{1984ApJ...278L...1N} 
data by \cite{1987Natur.329..519B}, who proposed an ancient planetary nebula (PN) origin. This 
interpretation was based on several arguments: (a)~the bipolar, toroidal morphology of the structure; 
(b)~its symmetric centering on the central star;  (c)~the large inferred total mass, which they argued 
is more consistent with a PN than with a nova remnant; (d)~the absence of a significant temperature 
gradient across the nebula; and (e)~kinematic measurements suggesting expansion velocities of 
only a few \kms, implying a timescale compatible with a PN origin. They excluded the possibility 
that the structure represents swept-up interstellar material produced by the nova
outburst, arguing that the nova shock would neither be energetic enough nor capable of generating such a 
large-scale structure. In particular, they noted that the \hi\ line width shows little evidence for velocities 
exceeding $\sim$5~\kms \citep{1989ApJ...344..805S}. 

\cite{1989ApJ...344..805S} describe \hi\  emission of the region, which has a morphology similar to the 
IRAS dust nebula. Their main conclusion was that, while it is plausible that the bipolar structure is a PN, 
there are several issues with this interpretation: (a)~asymmetry in the nova or/and PN outburst is 
required; (b)~the estimated age of $\sim100\,000$~years for the PN, which taking into account the theory 
of the isolated cooling white dwarf (WD), would require a higher luminosity for the WD primary than is 
observed in the case of GK Per; (c)~the complex structure of the interstellar medium around GK Per 
compared to the features in the IRAS and \hi\ images, so that it is inconclusive whether or not their 
origin is circumstellar.   

\cite{1989MNRAS.239..759H} presented $^{12}\mathrm{CO}$ observations of the same region, and 
concluded that at least part of this emission originates from interstellar cirrus, suggesting that the 
large-scale environment around GK~Per likely includes a contribution from unrelated interstellar material 
in addition to any circumstellar component. They argue that, while the $^{12}\mathrm{CO}$ emission in the 
southeast (SE) lobe correlates well with the far-infrared emission, it is completely absent from the 
northwest (NW) side. The estimated  $^{12}$CO mass (2\Msun) is inconsistent with that expected from 
formation of  a PN due to mass loss from the progenitor star \citep[e.g.,][]{2025A&A...694A.177V}. Furthermore, 
the $^{12}\mathrm{CO}$ velocities were no greater than a few \kms.

\cite{1994MNRAS.269..707S} reported a high-spatial-resolution study of the immediate environment of 
GK~Per in the $J = 2-1$ transition of $^{12}\mathrm{CO}$. Their observations revealed an apparent 
bipolar emission structure centered on the nova, with lobes expanding symmetrically at velocities of 
approximately 2.0--4.5~\kms, comparable to those measured by \cite{1989ApJ...344..805S}. The morphology 
of the CO emission broadly resembles the bipolar structure seen in the IRAS data. However, the detected 
features are located near the edge of the field of view, indicating that the CO nebula likely extends beyond 
the mapped area. Based on the observed velocity field, they favored an interpretation in terms of bipolar lobes
rather than a toroidal geometry. The inferred dimensions, expansion velocities, and total mass of the 
molecular nebula are consistent with those of PNe, with an estimated total hydrogen mass of $\sim0.05$\Msun, 
and an age of $\sim2.4\times10^{5}$~yr. On this basis, they concluded that GK~Per is surrounded by a 
fossil PN rather than a coincidental interstellar matter. However the observations of \cite{2022A&A...659A.109K} 
cast doubt on this conclusion. These authors analyzed archival Wide-field Infrared Explorer
\citep[WISE,][]{1996SPIE.2744..751S} data in conjunction with  new, more extended, CO observations in 
the $J = 1-0$ and $J = 2-1$ transitions. They concluded that, while the CO emission is most probably 
interstellar, the emission in WISE and in the optical range is most probably circumstellar.

The H$\alpha$ counterpart of the structure was discovered by \cite{1995ApJ...438..917T}. It shows the same 
bipolar morphology, but with a dominant ridge in the south-west (SW) that is not seen in the IRAS 
data. However, it corresponds to the region containing the brightest light echoes detected between 1901 
and 1902 \citep{1901ApJ....14..167R,1901ApJ....14..293R,1902ApJ....15..129R}. \cite{1995ApJ...438..917T} 
argued against a PN interpretation, noting that the temperature and inferred evolutionary state of the WD are 
inconsistent with the age required for a PN of this size. Instead, they proposed scenarios in which the nebula 
consists of accumulated material from (i) multiple previous nova eruptions or (ii) sustained mass loss 
during extended quiescent phases. This interpretation has credence as the nebula was already present 
at the time of the light echoes, implying a long-lived structure formed through repeated or prolonged 
mass ejection rather than a single event. On the other hand, \cite{1996A&A...306..547D} argue that 
the extended emission is physically related to the central binary, based on the symmetric morphology 
seen in the IRAS data. However, they assert that a more plausible explanation is that the secondary star 
evolved off the main sequence and expanded, leading to a phase of enhanced mass transfer 
onto the WD. This triggered a brief born-again asymptotic giant branch–like phase, accompanied 
by substantial mass loss, giving rise to the extended bipolar nebulosity.

All of the  above studies, including also the analysis by \cite{2004ApJ...600L..63B}, demonstrate the 
extensive debate,  and many questions relating to the nature of GK Per. In this paper we present wide 
field low resolution infrared spectral imagery recently obtained by the  Spectro-Photometer for the History 
of the Universe, Epoch of Reionization, and Ices Explorer 
\citep[SPHEREx,][and details therein]{2020SPIE11443E..0IC, 2025arXiv251102985B}
in order to address some of these questions. 

\begin{figure}
\includegraphics[width=0.465\textwidth, trim={0.01cm 0.01cm 0.01cm 0.01cm}, clip, keepaspectratio]{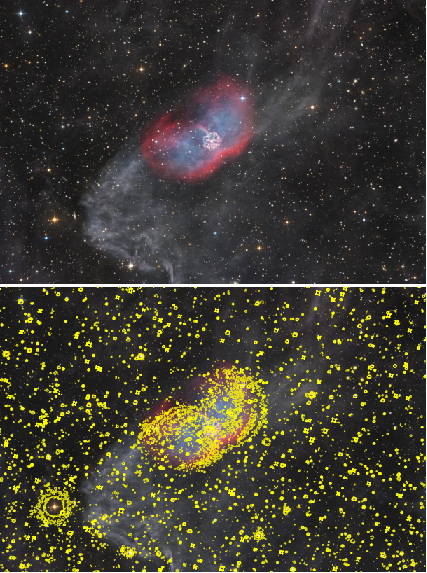}
\caption{\label{fig:DSC} The top panel shows a deep image of GK~Per and its surroundings obtained 
by the Deep Sky Collective (DSC\footnote{\url{https://ssr.app.astrobin.com/i/ocm8rv?r=0}}). 
The total exposure time was 265h 50m. Images were obtained in various filters. The per-filter exposure 
times were 13h 00m (Lum/clear), 14h 20m (R), 16h 15m (G), 15h 45m (B), 91h 55m (\oiii, 5 nm bandpass), 
14h 35m (H$\alpha$, 5~nm bandpass). For the color palette, the broadband is RGB with luminance being 
used for details and signal (the infrared cirrus hence appears grey). H$\alpha$ is red and \oiii{} is mapped 
to R and G, giving it a cyan-blueish look. The field of view is $50\arcmin \times 43\arcmin.$ North is up, and 
East is to the left. The bottom panel shows contours at surface brightness levels of 0.72, 1.05 and 
1.20 MJy/sr in the 4.6947~\micron{} H$_{2}$ emission line (discussed in Section~\ref{sec:obsresults}) 
observed by SPHEREx superposed on the optical image. }
\end{figure}

For comparison with the SPHEREx data, we utilize a previously unpublished optical image of the GK~Per 
region, shown in Figure~\ref{fig:DSC}. The wide field of view is needed in order to compare the morphology 
and characteristics  of the putative PN in the optical with those in the infrared.

 \section{Observations and Results}\label{sec:obsresults}

SPHEREx is a NASA space telescope that produces an all-sky map in 102 color bands covering the 
range 0.75--5.0~\micron{} It contains six linear Variable Filters (LVFs) arranged over six HgCdTe detector 
arrays and thus sequentially collects data in six wavelength bands, whose characteristics are summarized 
in Table~\ref{spherex}.

SPHEREx obtained 114 spectrophotometric images of GK~Per and its environs, covering all 102 of 
its wavelength bands. The exposure time for each image was 113.58 s. Emission was detected in 
$\sim$30 of the images. Because of the low spectral resolution (see  Table~\ref{spherex}), precise 
wavelengths of each emission feature cannot be determined from the spectra. However, the wavelengths 
of each of the bands in which emission was detected encompass sets of lines of the $v$=1--0 band and 
pure rotational  ($v$=0--0) band of molecular hydrogen (H$_2$). All of these detections are self-consistent 
with this identification, in the sense that all of these lines of H$_2$ might be reasonably expected to be 
present if some are present. The detected lines, numbering about 30 are listed in Table~\ref{spherex}. 
A few of the detections are blends of H$_2$ lines of different intrinsic strengths at nearby wavelengths 
that are not resolved by SPHEREx. Due to the weakness of the H$_2$ emission and thus the low 
signal-to-noise ratios on all but the strongest lines, not all of the lines in Table~\ref{spherex} are clearly 
detected in all 30 of the images at all locations in GK~Per where the stronger lines are detected. However, 
they are present in a sufficient number of locations that we regard all of these individual lines and 
blends as real detections, rather than artifacts. To our knowledge, these observations constitute the 
first detection of H$_{2}$ in GK~Per.
 
\begin{figure}
\includegraphics[width=0.485\textwidth, trim={0.01cm 0.01cm 0.01cm 0.01cm}, clip, keepaspectratio]{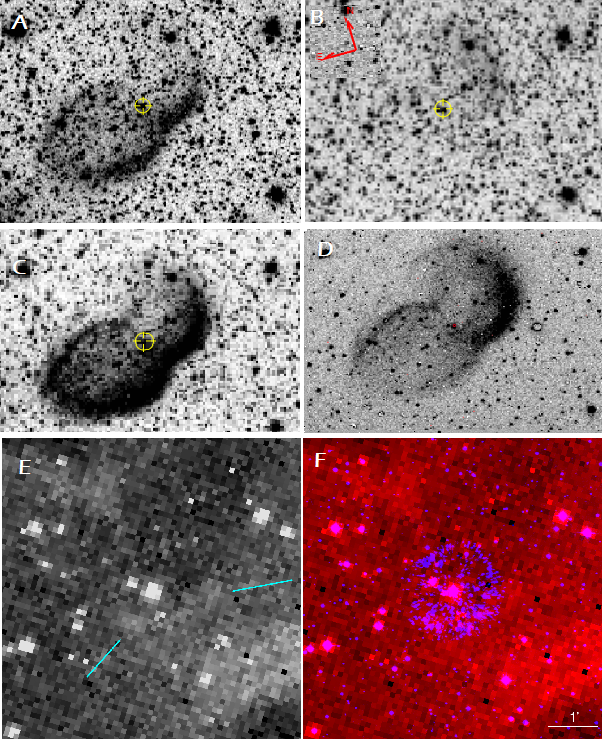}
\caption{SPHEREx images of GK~Per in the  1--0 $S$(1) 2.1218\micron{} (panel A), 
the combined 1--0 $S$(0) 2.2235~\micron{} and 2--1 $S$(1) 2.2477~\micron{}  (panel B),
1--0 $Q$(1) 2.4066 \micron{} (panel C) and 0--0 $S$(9) 4.6947 \micron{} lines (panel D). The first three afore-mentioned 
images are presented with the same color and stretch settings. Panels A through D have a field of view
$\approx 24\farcm0 \times 18\farcm5.$ The weakness of the combined 1--0 $S$(0) and 2--1 $S$(1) lines compared 
to that of 1--0 $S$(1) (2.1218 \micron{}, indicates that shock excitation
rather than UV-fluorescence is the excitation mechanism (see text). The 4.6947 \micron{} image (middle right) shows a 
filament of  H$_{2}$ emission about 44\arcsec\ south-west (PA = $145^\circ$) of the central star. This filament 
is magnified in panel E (North to the top) and shown between markers in cyan. Overlaid on this image, but now shown 
in red in panel F, is an H$\alpha$ image taken by us in 2023 showing the nova ejecta of 1901 in blue. The 
SW rim of the nova ejecta coincides with the filament over a large extent. The position of the filament was 
the site of intense X-ray and non-thermal radio emission. The size scale of 1\farcm0  
valid for panel E and F only is inset in the lower right of panel F.} 
\label{fig:imag1}
\end{figure}

Images of the 1--0 $S$(1) 2.1218 \micron{}, the combined 2--1 $S$(1) 2.2477 \micron{}
and 1--0 $S$(0) 2.2235~\micron{}, the 1--0 $Q$(1) 2.4066 \micron{}, and the 0--0 $S$(9) 4.6947 \micron{} 
lines are shown in Figure~\ref{fig:imag1}.  As can be seen by comparing these 
images with Figure~\ref{fig:DSC},  emission regions in the three brightest lines have the same 
orientation on the sky as the H$\alpha$ emission, have similar dimensions ($18\arcmin \times10\arcmin$) as 
the H$\alpha$ emission, and, like it, are strongest at the edges of the region.

\begin{figure*}
\begin{center}
\includegraphics[width=0.885\textwidth, trim={0.01cm 0.01cm 0.01cm 0.01cm}, clip, keepaspectratio]{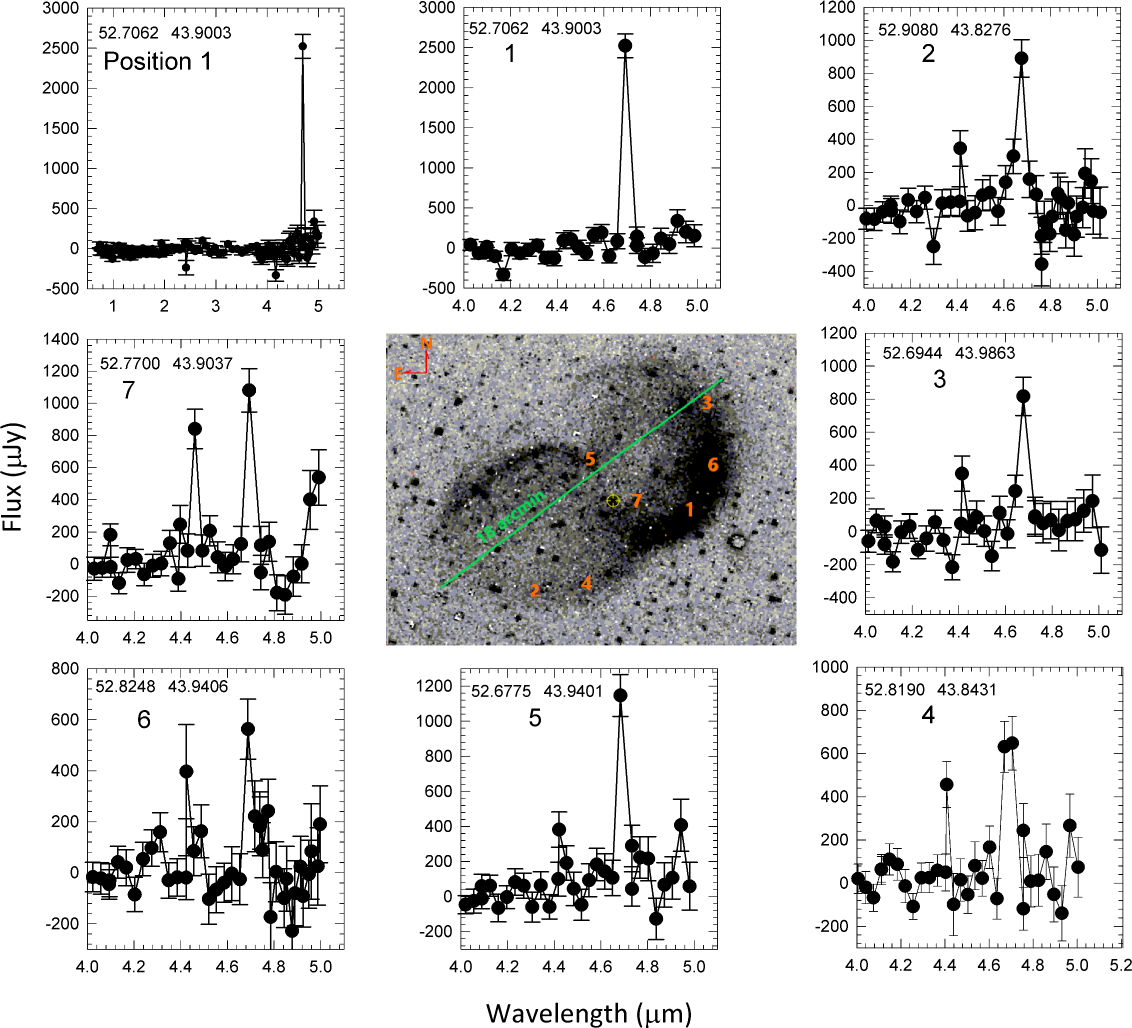}
\caption{SPHEREx spectra at selected positions on GK~Per whose image in the H$_{2}$ 0--0 $S$(9) 4.6947 \micron{}  line
is shown at center. Markers on the image indicate positions where spectra  are shown  here and whose 
spatial location on the nebula and corresponding RA and Dec (2000) positions (in degrees) are indicated 
at the top left in each panel.  The top left panel shows a representative spectrum covering the entire 
0.75--5 \micron{}  region measured at position 1; the rest of the spectra focus on the region covering 
the 0--0 $S$(9) 4.6947 \micron{} line which was the strongest line in the spectrum and which is seen at almost 
all the positions. More details are given in the text (Section~\ref{sec:disc-results}.)
The upward rise of some the spectra near 5.0 \micron{} is likely due to the 0--0 $S$(8) 5.0529 \micron{} line.} 
\label{fig:imag2}
\end{center}
\end{figure*}

We have extracted spectra at several locations in the nebula using the spectrophometry tool available
at the SPHEREx website\footnote{https://www.jpl.nasa.gov/missions/spherex/}.  Several 4 to 5~\micron{} segments 
of these spectra are presented in Figure~\ref{fig:imag2}, along with one spectrum covering the entire 
wavelength range. Each of the 4$-$5~\micron{} spectra  shows emission in the 0--0 $S$(9) line. As can be 
seen in central panel of the figure, that line is present over nearly the entire $18\arcmin \times10\arcmin$ nebula. 
The 0--0 $S$(10) 4.4096~\micron{} line is also present in most of these spectra, since emission is present 
in its color band.

In general, the intensity ratios of H$_{2}$ lines allow one to determine H$_{2}$ temperature and distinguish 
between excitation mechanisms for the molecule, in particular between collisional (e.g., shock) excitation 
and UV excitation. However, the inherent process by which SPHEREx data are recorded does not allow 
one to compare line strengths directly to use as diagnostics. Wavelength scanning in SPHEREx is 
done by nodding the telescope several times in the north-south direction, so that a star lies on different 
positions of the LVF (and hence at different wavelengths). Thus, for an extended source (such as the GK~Per 
nebula), there is a vertical wavelength gradient  between its north and south tips. Hence comparing 
he peak intensities of two lines from the same location on the nebula cannot be done directly because 
at that position the two lines could be shifted by different amounts  from their corresponding rest wavelengths.

Nevertheless, some basic conclusions can be reached. Both the large range of rotational levels that
are populated in the $\upsilon=0$ state, and the high intensity of the 0--0 $S$(9) line relative to the other 
pure rotational lines indicate collisions are the dominant excitation mechanism. This conclusion is 
supported by the obviously very low intensity ratio of the combined 1--0 $S$(0) and 2--1 $S$(1) lines at 2.224~\micron{} and
2.248~\micron{}, respectively, to that of the 1--0 $S$(1) line at 2.122~\micron, apparent from the examination
of panels A and B in Figure~\ref{fig:imag1}. Historically, the ratio of these two lines is the 
most common way of distinguishing between collisional and UV excitation
as discussed in \citet[][see Figure 2]{1987ApJ...322..412B} or \citet[][]{1987ApJ...318L..73G}. 

Going beyond this basic level of analysis of relative line intensities is highly problematic. For example, 
in  Figure~\ref{fig:imag2} the full spectrum at Position 1 is totally dominated by the 0--0 $S$(9) line; all 
other lines appear weaker than it by at least an order of magnitude. We believe there is no reasonable 
physical explanation for this. Very high extinction by dust could account for some of the anomaly, but not 
all of it and, in addition, no evidence for high extinction by dust exists. Moreover, the nearby 
$S$(10) 4.4096 $\mu$m line, which should be roughly one-third the strength of the $S$(9) line (as the 
ratio of ortho-to-para hydrogen is generally 3:1 in statistical equilibrium) and should suffer nearly the 
same extinction by dust, is not detected. We conclude that determination of line strengths 
requires a level of data analysis beyond the scope of this paper, as well as a better understanding 
by us of the details of the performance of SPHEREx.

\begin{deluxetable*}{cccccccc}
\tablewidth{0pt}
\tablecaption{Observational details and list of detected lines\label{spherex}}
\tablehead{}
\startdata
SPHEREX & $\lambda$& $R^a$ &SPHEREx& Date & H$_2$   & $\lambda$ (vac)& $E_2$\\
band    & range (\micron{}& & ID   & MJD  & line  & (\micron{} & (K)$^b$ \\ \hline
D6 & 4.42--5.00& 128&2025W35\_1A\_0217\_2 &60913.035283 & 0--0 $S$(10) &4.4096  &11940\\
$''$ & & &2025W35\_1A\_0518\_2 &60914.664914 & 0--0 $S$(9) &4.6947& 10263\\
$''$ &  &&2025W35\_1A\_0518\_3 &60914.666407 & 0--0 $S$(9) &4.6947& 10263 \\
D5 & 3.82--4.42&112 &2025W34\_1B\_0174\_3 &60905.904597 & 0--0 $S$(11) & 4.1810&13703 \\
$''$ & & &2025W34\_1B\_0174\_4 &60905.906090 & 0--0 $S$(10) & 4.4096& 11940\\
$''$ & & &2025W34\_1B\_0187\_2 &60905.971393 &  0--0 $S$(10) & 4.4096& 11940\\
$''$ & & & 2025W33\_1B\_0598\_4 & 60911.679808&  0--0 $S$(12) &3.9947& 15549 \\
$''$ & & & 2025W33\_1B\_0393\_2 &60910.589958 & 0--0 $S$(13) &  3.8464 & 17458\\
$''$ & & & 2025W33\_1B\_0393\_3 &60910.591468 & 0--0 $S$(13) & 3.8464 & 17458 \\
$''$ & & & 2025W33\_1B\_0598\_2 & 60911.676816& 0--0 $S$(13) &  3.8464 &17458 \\
D4	&2.42–3.82	&35		&	2025W34\_1B\_0035\_2	&	60905.152963	&	1--0 $O$(7)	&	3.8075	&	8365	\\
$''$ & & & 2025W34\_1B\_0102\_1  & 60905.492522& 1--0 O(6) & 3.5007  & 7485 \\
$''$	&		&		&	2025W34\_1B\_0102\_1	&	60905.492509	&	1--0 $O$(6)	&	3.5007	&	7485	\\
$''$ & & & 2025W34\_1B\_0048\_2 &60905.2219708& 1--0 $O$(5) & 3.2350 & 6956 \\
$''$	&		&		&	2025W34\_1B\_0048\_1	&	60905.220416	&	1--0 $O$(4)	&	3.0039	&	6471	\\
$''$ & & & 2025W34\_1B\_0062\_1  & 60905.289948& 1--0 O(4)  &3.0039  & 6471 \\
$''$	&		&		&	2025W34\_1B\_0048\_2	&	60905.221908	&	1--0 $O$(5)	&	3.2350	&	6956	\\
$''$	&		&		&	2025W34\_1B\_0062\_1	&	60905.289916	&	1--0 $O$(4)	&	3.0039	&	6471	\\
$''$	&		&		&	2025W33\_2C\_0028\_2	&	60901.619552	&	1--0 $O$(4)	&	3.0039	&	6471	\\
$''$	&		&		&	2025W33\_2C\_0014\_3	&	60901.554479	&	1--0 $O$(3)	&	2.8025	&	6149	\\
$''$	&		&		&	2025W33\_2C\_0014\_4	&	60901.555972	&	1--0 $O$(3)	&	2.8025	&	6149	\\
$''$	&		&		&	2025W33\_2C\_0014\_1	&	60901.551506	&	1--0 $O$(2)	&	2.6269	&	5987	\\
$''$	&		&		&	2025W33\_2C\_0014\_2	&	60901.552987	&	1--0 $O$(2)	&	2.6269	&	5987	\\
$''$ & & & 2025W33\_1B\_0652\_4  & 60901.488465& 1--0 $Q$(7) & 2.5001  & 10341 \\
$''$ & & & 2025W33\_1B\_0652\_3  & 60901.488465 & 1--0 $Q$(6) & 2.4756  & 9286 \\
$''$ & & & 2025W33\_1B\_0652\_3  &60901.488465& 1--0 $Q$(4) to $Q$(7) & 2.4548 & 8365 \\
$''$ & & & 2025W33\_1B\_0652\_2  & 60901.486973& 1--0 $Q$(1) to $Q$(4)  & 2.4375  & 7586 \\
D3	&1.64-2.42	&	41	&	2025W34\_2B\_0490\_2	&	60911.134133	&	1-0 $Q$(1) to Q(4)&		2.4066	&	6149	\\
$''$	&		&		&	2025W34\_2B\_0443\_4	&	60910.864888	&	1-0 $S$(0)	&	2.2235	&	6471	\\
$''$	&		&		&	2025W35\_1A\_0187\_1	&	60912.897232	&	1-0 $S$(0)	&	2.2235	&	6471	\\
$''$	&		&		&	2025W34\_2B\_0443\_3	&	60910.863395	&	1-0 $S$(1)	&	2.1218	&	6956	\\
$''$	&		&		&	2025W34\_2B\_0304\_1	&	60910.113145	&	1-0 $S$(2)	&	2.0338	&	7584	\\
$''$	&		&		&	2025W34\_2B\_0443\_1	&	60910.860411	&	1-0 $S$(2)	&	2.0338	&	7584	\\
$''$	&		&		&	2025W34\_2B\_0443\_2	&	60910.861903	&	1-0 $S$(1)	&	2.0338	&	6956	\\
$''$	&		&		&	2025W35\_1A\_0518\_3	&	60914.666407	&	1-0 $S$(2), S(3)	&	2.0338	& 7584	\\
$''$	&		&		&	2025W37\_2A\_0034\_1	&	60929.610165	&	1-0 $S$(2)	&	2.0338	&	7584	\\
$''$	&		&		&	2025W35\_1A\_0518\_2	&	60914.664914	&	1-0 $S$(3)	&	1.9576	&	8365	\\
$''$	&		&		&	2025W35\_1A\_0518\_1	&	60914.663422	&	1-0 $S$(4)	&	1.8920	&	9286	\\
$''$	&		&		&	2025W35\_1A\_0441\_4	&	60914.260211	&	1-0 $S$(5)	&	1.8358	&	10341	\\
$''$	&		&		&	2025W35\_1A\_0441\_3	&	60914.258719	&	1-0 $S$(6)	&	1.7880	&	11522	\\
$''$	&		&		&	2025W35\_1A\_0441\_1	&	60914.255735	&	1-0 $S$(7), S(8)	&	1.7480	&12817	\\
$''$	&		&		&	2025W35\_1A\_0441\_2	&	60914.257227	&	1-0 $S$(7)	&	1.7480	&	11522	\\
\enddata
\tablecomments{$^a$Resolution $R=\lambda/\Delta{\lambda}$. 
$^b$Energy of upper level.
All observations were 113.58~sec. The SPHEREx IDs were taken from the IRSA/IPAC site}
\end{deluxetable*}

A faint arc of emission is evident in the 4.69947 \micron{} image containing the H$_{2}$ 0--0 $S$(9) line, which
coincides  well with the SW edge of the nova ejecta. This may be seen by overlaying a H$\alpha$ image 
taken by us in 2023 October with the Isaac Newton Telescope (see Figure~\ref{fig:imag1}). The H$_{2}$ arc 
also coincides, over a large part,  with the intense X-ray arc \citep{2005ApJ...627..933B} and with the radio arc from 
synchrotron emission \citep{1989ApJ...344..805S,1995MNRAS.276..353S} found at the same position. 
A composite optical/radio/X-ray picture showing this arc is available at \url{https://chandra.harvard.edu/photo/2015/gkper/. }
Likely this arc is cool H$_{2}$-bearing material between the forward and reverse shock as 
the nova ejecta encountered the ambient medium. The presence of such cool material was argued 
for by \cite{2005ApJ...627..933B}, based on the greater neutral H absorption seen in  the X-ray component 
associated with  the reverse shock vis-a-vis a lesser neutral H absorption in the forward shock. This 
additional absorption indicated a neutral H zone lay between the shocks \citep{2005ApJ...627..933B}. 
Theoretically, such a zone between the forward and reverse shocks is predicted, both in novae 
and supernovae, as their ejecta propagate into their  surrounding medium \citep[e.g.,][]{1996ApJ...461..993F,2017MNRAS.469.1314D}.
This cool zone is precisely where molecules and dust are proposed to form \citep{2017MNRAS.469.1314D}, 
so the presence of H$_{2}$ in the arc is consistent. This result is a rare synthesis of theory, and optical, 
radio, X-ray, and infrared observations. 

\section{Discussion}\label{sec:disc-results}

The discovery of the H$_{2}$ envelope around GK~Per may favor the classification 
of its large nebulosity as a PN. PNe, and bipolar PNe specifically, are often known to have such H$_{2}$ 
shells  \citep{1996ApJ...462..777K, 2003PASP..115..170S, 2017MNRAS.470.3707R}.
For GK~Per, the mass of the  putative PN, based on the H$\alpha$ nebulosity seen in Figure~\ref{fig:DSC},
can be estimated by assuming the  two-dimensional $18\arcmin\times10\arcmin$ PN to be approximately 
an ellipsoid in three dimensions with semi-major axes $a,b,c = 9\arcmin,5\arcmin,5\arcmin$. For an 
assumed distance of 470~pc, assuming that  its hydrogen is fully ionized and that the electron 
density $n_e=40$~cm$^{-3}$ (this value is justified below),  the mass lies in the range 0.28--0.38\Msun, 
assuming a mean value of the filling factor between 0.3 to 0.4  \citep{1994A&A...284..248B}. This is 
consistent with the range of PNe masses, 0.1--1.0\Msun\ \citep{1984A&A...130...91P}. 

However, there are a few inconsistencies with the PN scenario. A bipolar shape invariably needs a 
dense equatorial constriction (e.g., a torus) to shape the nebula, but clear evidence for such a torus 
is not seen in the images here. Such a torus or  equatorial constriction is expected to be dense and 
hence the likely site of H$_{2}$ and dust 
\citep[e.g., the hourglass nebula NGC 2346;][]{1986BAAS...18.1054G,2017MNRAS.470.3707R}.
The velocity kinematics across the field also appear to be on the  lower side, from the limited radio 
studies (see Section~\ref{intro}), compared to typical PNe expansion velocities $V_{\rm exp}$ of
10--30\kms, although very old PNe can have lower $V_{\rm exp}$ values \citep{1974ApJ...193..197B}.  
Observational tests for the  bipolarity of the nebula can be done by measurements of $V_{\rm exp}$ at 
the waist and poles. Bipolar flows generally follow a $V\propto{r}$ law (homologous flow) and thus 
the waist expands much more slowly than the poles.

Other arguments against a PN interpretation arise from the timescales involved in the evolution of a 
0.82 and a 1.0~\Msun{} star \citep[e.g.,][]{2016A&A...588A..25M}. These tracks were selected because 
the WD in GK~Per has a mass between $\simeq 0.8$ to 1.0~\Msun{} \citep{1986ApJ...300..788C, 
2002MNRAS.329..597M, 2021MNRAS.507.5805A}. 
From these tracks we note, that taking 30,000~K as the lower end of the temperature 
range for the central stars of PNe (CSPNe), PN should form within a few years.
The tracks also indicate that the time to reach WD luminosities 
of 10$^{-2}$ to 10$^{-3}$ \Lsun{} at which conventional nova models are 
generated \citep{1998ApJ...494..680J, 2024ApJ...962..191S, 2025ApJ...982...89S} is more than a
million years. Since the general lifespan of a PN is between $\simeq 10,000$ to 30,000~yrs, the PN 
would have faded below detection level by the time the nova erupted in 1901. Hence, it should
not be possible to see the PN and the nova remnant at the same time. Further, the mass of the 
present WD in GK~Per is higher than the known mass of any CSPN, irrespective of whether the
CSPN is a single star or is in a binary system 
\citep[see Figure 13 of][for a histogram of CSPNE masses]{2020A&A...640A..10W}.
The model WD luminosities used, between 10$^{-2}$ to 10$^{-3}$~\Lsun{} that are needed to initiate 
and drive the nova eruption, could not have been that of a CSPN because there are no observed CSPNe 
with luminosities $\ltsimeq 1$~\Lsun. We have determined this by plotting the archival data within Figure~5 
(histogram of CSPNe luminosities) in \citet{2020A&A...640A..10W}. From all these arguments
the nebulosity cannot be a conventional PN.

We have also considered whether the PN could be a nova super-remnant (NSR) created by repeated 
eruptions of a recurrent nova sweeping out the circumstellar material. NSR’s were first discovered around 
M31N 2008-12a \citep{2019Natur.565..460D} and subsequently around the Galactic RNe KT~Eridani
\citep{2024MNRAS.529..224S, 2024MNRAS.529..236H}, T~CrB \citep{2024ApJ...977L..48S} and
RS~Ophiuchi \citep{2025AJ....170...56S}. However, these NSRs have diameters ranging from 
$\sim 30$ to 130~pc whereas the H$\alpha$ nebulosity here is much smaller. Based on its angular size, the 
semi-minor and major axes are $\sim 0.7$ and 1.2~pc respectively.

Thus, as an alternative to the PN scenario, we consider whether the nebulosity is actually material that 
pre-existed the 1901 eruption of GK~Per, which was  thereafter ionized by the nova outburst and is 
presently in the recombination phase. There is substantial material around GK~Per, as evidenced in 
\hi\ and CO mapping studies, IRAS and WISE imagery, and the detection of a light echo 
(Section~\ref{intro}). The H$\alpha$ gas in  the nebulosity is  certainly at low density, because imaging 
of the jet by \citet[][see their Figure~12]{2012AJ....143..143S} also showed parts of the nebulosity 
emitting in H$\alpha$,  and it is clearly seen that the surface brightness of the jet is substantially
greater than that of the PN nebulosity. Since the  density of the jet was estimated to be 20-40~cm$^{-3}$ 
from the \sii\ lines \citep{2012AJ....143..143S}, assuming $n_{\rm e} =40 $~cm$^{-3}$  as the upper 
limit for the nebulosity is justified. The Case B recombination time for ionized hydrogen with
density $n_{\rm e}$ and temperature $T_{\rm e}$  is $(\alpha(T_{\rm e}) n_{\rm e})^{-1}$. Here, 
the recombination rate coefficient $\alpha(T_{\rm e}) =  2.6\times10^{-13}$~cm$^3$~s$^{-1}$ 
for an assumed $T_{\rm e} = 10000$~K under Case B conditions. The recombination  time is 
hence  $\sim3000$~years,  and could  explain why the nebulosity remains visible today. 

An apparent inconsistency in this scenario is that the secondary, if evolving as a single star, would 
take $\gtsimeq 1$~Gyr to evolve to its present K2 IV state. In contrast, a 0.9~\Msun{} WD would take
$(5 - 8) \times 10^{7}$~yr to evolve from the main-sequence to the WD stage \citep{1996A&A...306..547D}.
This mismatch between time scales can possibly be resolved if there has been mass exchanges between 
the two components in the past, through Roche Lobe overflow \citep{1996A&A...306..547D}. 
This may explain why the K type secondary, whose mass has variously been estimated between 
0.25 to 0.48~\Msun{} is under massive for its spectral type  \citep{1986ApJ...300..788C, 2021MNRAS.507.5805A}.
Similar systems do exist, such as UCAC2~46706450 which has a WD with a K subgiant secondary \citep{2020A&A...642A.228W}.
Furthermore, surveys of \citet{2016MNRAS.463.2125P} have established a large group of main 
sequence FGK stars which show excess UV flux, and hence likely have a WD companions \citep{2020A&A...642A.228W}.

GK Per and its environment remain partially understood and the more we discover about it, the more
we realize how incomplete is our understanding of it. There is a need for a spatio-kinematic study of the 
nebulosity to further test its classification as a PN, and thence understand its relation to the 1901 outburst.

\facilities{SPHEREx, INT:Newton}

\begin{acknowledgments}
The authors thank the referee for their insightful and thoughtful comments that improved the manuscript.
This publication makes use of data products from the Spectro-Photometer for the History of the Universe, 
Epoch of Reionization and Ices Explorer (SPHEREx), which is a joint project of the Jet Propulsion 
Laboratory and the California Institute of Technology, and is funded by the National Aeronautics and Space Administration. 
SPHEREx data used in this manuscript can be found at 10.26131/IRSA629 SPHEREx Quick Release 
Spectral Images, https://doi.org/10.26131/IRSA629.
We thank the many amateur astronomers who contributed to the striking Deep Sky 
Collective\footnote{https://deepskycollective.com/home} image, and Tim Schaeffer for permission 
to reproduce it here. We thank Marcelo Miller Bertolami for kindly providing the evolutionary tracks for the 1~\Msun{} star.
SS acknowledges partial support from a NASA Emerging Worlds grant to ASU (80NSSC22K0361) as well as support from 
his ASU Regents’ Professorship and JWST/NASA.  CEW acknowledges partial support from NASA JWST-06866.022-A.
VJ and DPKB acknowledge support from PRL, Ahmedabad and the Department of Space, Government of India.
\end{acknowledgments}

\bibliography{sample701-6}{}
\bibliographystyle{aasjournalv7}

\end{document}